\documentstyle[twoside,fleqn,espcrc2]{article}

\newcommand{\AmS}{{\protect\the\textfont2
  A\kern-.1667em\lower.5ex\hbox{M}\kern-.125emS}}

\title{The two-phase
issue in the $O(n)$ non-linear $\sigma$-model: a Monte Carlo study}

\author{B. All\'es\address{Dipartimento di Fisica and INFN,
        Piazza Torricelli 2, 56100 Pisa, Italy}\thanks{Speaker 
        at the conference.}
        A. Buonanno$^{\rm a}$ and G. Cella$^{\rm a}$
        }
       
\begin{document}

\begin{abstract}
We have performed a high statistics Monte Carlo simulation to 
investigate whether the
two-dimensional $O(n)$ non-linear sigma models are asymptotically free 
or they show a Kosterlitz-Thouless-like phase transition. We have
calculated the mass gap and the magnetic susceptibility 
in the $O(8)$ model with standard action and the $O(3)$ model with
Symanzik action. Our results for $O(8)$ 
support the asymptotic freedom scenario.
\end{abstract}

\maketitle

\section{INTRODUCTION}

The 2-dimensional 
$O(n)$ non-linear $\sigma$-model is defined by the action
\begin{equation}
S = {\beta \over {2}} 
\int {\rm d}^2 x \left(\partial_{\mu} {\vec \phi}\right)^2
\end{equation}
together with the condition ${\vec \phi}(x)^2 = 1$ for all spacetime
points $x$. In this equation $\beta$ is the inverse of the 
bare coupling constant.

Perturbation theory $(PT)$ 
predicts that this model is asymptotically free
for $n \geq 3$. In particular the exponential 
correlation length $\xi$ on the
lattice must scale as
\begin{equation}
\xi = C_{\xi} \left({1\over {2 \pi \beta \Delta}}\right)^{\Delta}
e^{2 \pi \beta \Delta} \left( 1 + \sum_{k=1}^{\infty}
{a_k \over \beta^k} \right)
\end{equation}
where $a_k$ are the corrections to universal scaling. Here
$\Delta\equiv 1/(n-2)$. $C_{\xi}$ is a non-perturbative constant
which for the standard action equals
\cite{hnm}
\begin{equation}
C_{\xi}=\left({e \over 8}\right)^{\Delta} \Gamma(1+\Delta) 2^{-5/2}
\exp \left(-{{ \pi \Delta} \over {2}}\right).
\end{equation}
We define the magnetic susceptibility $\chi$ as the 
two-point correlation
function at zero momentum. It scales as
\begin{equation}
\chi = C_{\chi} \left({1 \over {2 \pi \beta \Delta}}\right)^{\Delta
(n+1)} e^{4 \pi \beta \Delta} \left( 1 + 
\sum_{k=1}^{\infty} {{b_k} \over {\beta^k}} \right)
\end{equation}
where again $C_{\chi}$ is a non-perturbative constant. From equations 
(2) and (4) we conclude that in $PT$ the ratio
\begin{equation}
R_{PT} \equiv {\chi \over {\xi^2}} 
(2 \pi \beta \Delta)^{{n-1} \over {n-2}} 
\left( 1 + \sum_{k=1}^{\infty} {{d_k} \over {\beta^k}} \right)
\end{equation}
tends to $C_\chi/C_\xi^2$ 
as we approach the continuum limit, $\beta \rightarrow
\infty$. The corrections to asymptotic scaling $d_k$ depend on $\{a_k\}$ 
and $\{b_k\}$.

In a series of papers \cite{seiler}
another scenario has been put forward for the
model defined in (1). Under reasonable hypothesis the authors prove
that there is no mass gap and that this model
must undergo a Kosterlitz-Thouless-like ($KT$) 
phase transition at finite
beta, $\beta_{KT}$. This implies that the ratio
\begin{equation}
R_{KT} \equiv {\chi \over {\xi^{2 -\eta}}} 
\end{equation}
should be constant as one approaches $\beta_{KT}$ from below.
Here $\eta$ is a critical exponent. For the $O(2)$ model this exponent
is $\eta=1/4$. In \cite{summer95} the authors show that the
$O(3)$ model with the standard action on the lattice and $\eta=1/4$ 
gives a constant for $R_{KT}$ while the data for $R_{PT}$ 
displays a clear drop.

Here we will show a progress report 
from an extensive simulation performed on
the $O(8)$ model with standard action and the $O(3)$
model with the tree-level improved Symanzik action \cite{sym}. 
If the constancy of $R_{KT}$ for the $O(3)$ model is a genuine
physical effect, then also for the Symanzik action we should see such
a behaviour. The full account of our results with better statistics
and using more corrections to asymptotic scaling can be found
in \cite{we}.

\begin{figure}[htb]
\vspace{4.5cm}
\includegraphics{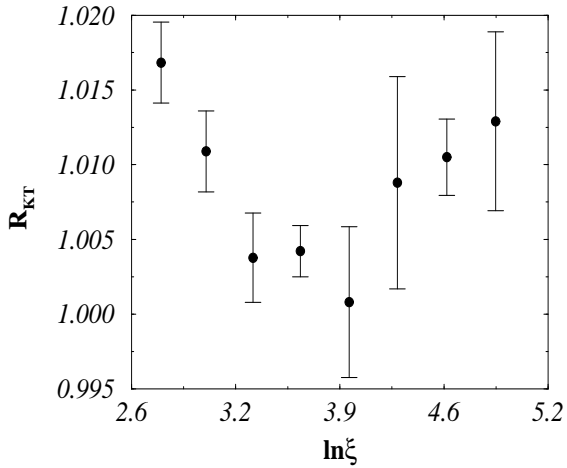} 
\null\vskip 0.3cm
\caption{The ratio $R_{KT}$ for the $O(3)$ model with Symanzik action.}
\end{figure}

\begin{figure}[htb]
\vspace{4.5cm}
\includegraphics{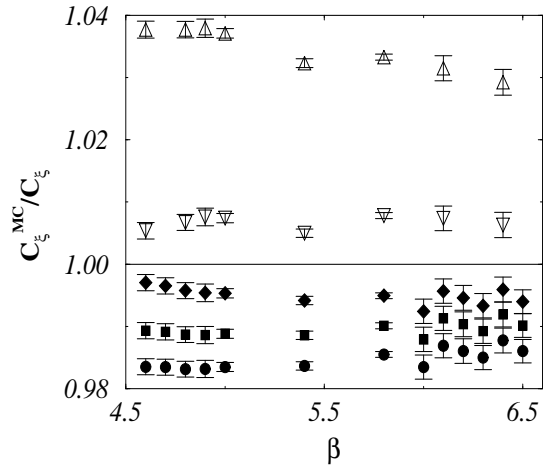} 
\null\vskip 0.3cm
\caption{The non-perturbative constant for the correlation length.
The 2, 3 and 4-loop results correspond to black circles, squares and
diamonds respectively. The 2 and 3-loop results in the energy scheme
are the up and down triangles.}
\end{figure}

\section{SIMULATIONS}

In our simulations we have used the Wolff algorithm \cite{wbc}
for the updatings
as well as improved estimators to measure the correlation length and
magnetic susceptibility. We have performed several millions of
measurements for both quantities and verified the absence of
autocorrelations.

To calculate the correlation length we have measured the second moment 
$\xi^{(2)}$. The ratio $\xi^{(2)}/\xi$ is less than few parts 
per mille, so within our statistical errors, we can use the formulae
(2-3).

We have chosen large enough lattice sizes $L$ 
to keep finite-size effects under
control. The ratio $L/\xi$ is $7-10$. We have checked that these 
effects are few parts per mille. 

The largest systematic error comes from the deviation from universal
scaling of our data. 
These corrections are known up to 4 loops for the standard
action and up to 3 loops for the Symanzik action \cite{trevescp}.

\section{RESULTS}


\subsection{The $O(3)$ model}

In figure 1 we show the results for $R_{KT}$ in the $O(3)$ model
with Symanzik action. They have better statistics than those
of reference \cite{summer95}. 
In constrast with \cite{summer95} our data are not constant.

We do not show here (see \cite{we}) 
the data for $R_{PT}$. Again it is
not constant although it displays better scaling than for the
standard action \cite{summer95}. The fits for $C_\xi$ and $C_\chi$ agree
with the prediction (3) and large-$n$ calculations within $15-20\%$.

Assuming finite-size scaling, it has been shown that this model 
presents asymptotic scaling 
starting from $\xi \approx 10^5$ \cite{kimcp}

\subsection{The $O(8)$ model}

In figure 2 we show the ratio between the non-perturbative constant 
$C_{\xi}^{\rm MC}$ as computed from our Monte Carlo data and the
prediction (3) which for the $O(8)$ model is $C_{\xi}=0.10544$. If $PT$
is correct and asymptotic scaling holds, this ratio should be equal to
1 (up to $\sim$0.1 per mille because we measured the second moment 
$\xi^{(2)}$).

\begin{figure}[htb]
\vspace{4.5cm}
\includegraphics{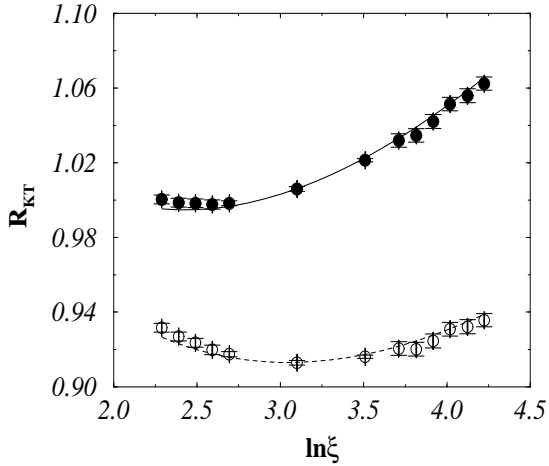} 
\null\vskip 0.3cm
\caption{The ratio $R_{KT}$ for the $O(8)$ model. Black and white circles
represent $\eta=0.25$ and $\eta=0.22$ respectively. The lines are the 
$PT$ predictions for these ratios.}
\end{figure}

\begin{figure}[ht]
\vspace{4.5cm}
\includegraphics{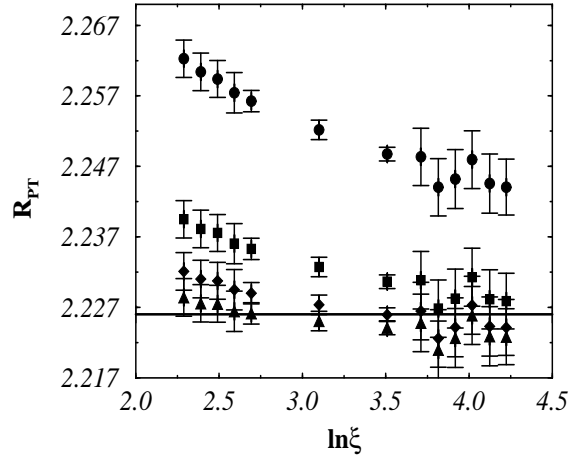} 
\null\vskip 0.3cm
\caption{The ratio $R_{PT}$ for the $O(8)$ model. The 
successive orders correspond to circles, squares, 
diamonds and triangles respectively.}
\end{figure}

We show the data as obtained from the 2, 3 and 4-loop
approximation in eq. (2). The
data in the scheme of the energy \cite{parisi} at 2 and 3-loop
are also shown (we have used the
energy measurements of reference \cite{mendes}). All data seem to
converge towards the $PT$ prediction. A careful analysis of the next
coefficients in the $1/n$ expansion suggests that further corrections
should have small effects. The 4-loop data in this figure agree 
with (3) within 0.5\%.

In figure 3 the data for $R_{KT}$ are shown. The data are far
from constant. We show the results for two values of $\eta$: $\eta=0.25$
is the upper set of data (black circles) and
$\eta=0.22$ is the lower set (white circles).

The solid and dashed lines are the $PT$ predictions for $R_{KT}$
assuming $C_{\chi}=0.103$ (this is the value obtained from a best fit
performed on our data for $\chi$; it agrees with large-$n$ estimates 
within 1\%) 
and the prediction (3) for
$C_{\xi}$. We see that not only the curves are not constant but also
that $PT$ explains well its non-constancy.

In figure 4 we show the data for $R_{PT}$. The upper set of data
is the lowest order prediction. Further corrections 
stabilize the result. The result clearly
converges to a constant. Physical scaling is well reached at the
largest correlation lengths. The
corrections to universal scaling converge surprisingly well.

The solid horizontal line is the $PT$ prediction for the constant by
using $C_{\chi}=0.103$ and the prediction (3) for the
correlation length. We conclude that our data are in fair agreement
with $PT$.

In conclusion our data do not support either $KT$ or $PT$ 
for the $O(3)$ model
but they show clear agreement with $PT$ for the $O(8)$ model. In the 
$KT$ scenario one should explain why $PT$ works so well for large $n$
and large ratios $L/\xi$.

\vskip 7mm

This work has benefited from many 
stimulating conversations with Andrea Pelissetto.

\end{document}